\magnification=\magstep1
\font\gross=cmbx12 scaled \magstep0

\font\Gross=cmr12 scaled \magstep2
\font\Mittel=cmr12 scaled \magstep0
\overfullrule=0pt
\def\Bbb#1{{\bf #1}}
\def\blacksquare{\sqcup\!\!\!\!\sqcap}
\baselineskip=12pt
\def\up{\uparrow}
\def\down{\downarrow}
\def\k{{\bf k}}

\def\p{{\bf p}}
\def\q{{\bf q}}

\def\ts{\textstyle}
\def\ds{\displaystyle}

\def\la{\langle}
\def\ra{\rangle}
\def\pro{\mathop\Pi}
\def\1cm{\hskip 1cm}

\def\vep{\varepsilon}

\def\wk{\sqrt{\kappa}}

{
\nopagenumbers
\baselineskip=14pt
$$ $$
\vskip 3cm
\centerline{\Gross The Global Minimum of the Effective Potential }
\smallskip
\centerline{\Gross of the Many-Electron System with
   Delta-Interaction}
\bigskip
\bigskip
\centerline{by}
\bigskip
\bigskip
\centerline{{\Mittel Detlef Lehmann}\footnote{$\phantom{}^*$}{ e-mail: 
 lehmann@math.tu-berlin.de}}
\centerline{\Mittel Technische Universit\"at Berlin}
\centerline{\Mittel Fachbereich Mathematik Ma 7-2} 
\centerline{\Mittel Sta{\ss}e des 17. Juni 136}
\centerline{\Mittel D-10623 Berlin, GERMANY}
\vskip 4cm 
\noindent{\bf Abstract:}  We prove that the global minimum 
of the real part 
 of the full effective 
 potential of the many electron system with attractive 
 delta interaction is in 
 fact given by the BCS mean field configuration. This is a 
 consequence of 
 a simple bound
 which is obtained by applying Hadamard's inequality to 
the functional 
 determinant.  The second order Taylor expansion around the minimum 
is computed. 
\bigskip
\bigskip
\vfill
\eject  }
\baselineskip=14pt

\noindent{\gross I. Introduction and Results} 
\bigskip
In this article we consider the effective potential $V$ of the  nonrelativistic 
 many electron system  with attractive delta function interaction. In 
 momentum space, it reads 
$$\eqalignno{ V(\{\phi_q\})&=\sum_q |\phi_q|^2
  -\log{ \det\left[\matrix{ Id & {ig\over 
  \sqrt{\beta L^d} }\, 
 C \phi^* \cr {ig\over\sqrt{\beta L^d} }\, 
   \bar C\phi & Id \cr}\right]  } & (1) \cr
 & \cr
 &=\sum_q |\phi_q|^2-\ts\log{\det\left[Id+{g^2\over\beta L^d}\,\bar C\phi C\phi^*
   \right]}  \cr}$$
where $g^2=\lambda>0$ is an attractive coupling, 
 $C=(\delta_{k,p}C_k)_{k,p}$ is a diagonal matrix with entries 
 $C_k={1\over ik_0-\vep_\k+\mu}$, $\vep_\k=\vep_{-\k}$ being the 
single particle 
 energy momentum relation  and $\mu$ denoting 
 the chemical potential. In the following we set $e_\k=\vep_\k-\mu$ such that the
Fermi surface is given by $e_\k=0$. 
 Furthermore $\phi$ is a short notation for the matrix 
 $(\phi_{k-p})_{k,p}$. The momenta $k=(k_0,\k),p=(p_0,\p)$ range 
 over some finite subset 
of ${\pi\over\beta}(2\Bbb Z+1)\times \left({2\pi\over L}\Bbb Z\right)^d$ if the 
system is kept in finite volume $[0,L]^d$ and at some small but positive 
 temperature $T={1\over \beta}>0$.  To be 
specific, choose 
$$\ts  k,p\in M_\nu:=\left\{ (k_0,\k)\in 
  {\pi\over\beta}(2\Bbb Z+1)\times 
 \left({2\pi\over L}\Bbb Z\right)^d\>\bigr|\> |e_\k|
  \le 1,\; |k_0|\le  \nu\right\} \eqno (2)$$
where $1<\!\!<\nu<\infty$ is some cuttoff. The momenta $q=(q_0,\q)$ 
 are given by $q\in\{k-p\>|\>k,p\in M_\nu\}$.  
\par
It is widely believed that the global minimum of the real part 
of $V$ (in general, the determinant in (1) is complex) is given by 
the mean field configuration $\phi_x=const$ in coordinate space or 
$\phi_q\sim\delta_{q,0}$ in momentum space [1,2,3]. Although this is 
suggested by diagrammatic arguments [4], there was, to the authors 
 knowledge, no rigorous proof of that. In Theorem 1 below we show 
that this result can be obtained by applying Hadamard's 
inequality (the absolute value of the determinant of $n$ column 
vectors is less than the volume of the cube spanned by the $n$ 
vectors) to the determinant in (1). 
\par
More precisely, all global minima of the real part 
$${\rm Re}V(\{\phi_q\})=\sum_q |\phi_q|^2-\log{\left| \det\left[\matrix{ Id & {ig\over 
  \sqrt{\beta L^d} }\, 
 C \phi^* \cr {ig\over\sqrt{\beta L^d} }\, \bar C\phi & Id \cr}\right]\right|  }
 \eqno(3)$$
of $V$ are given by 
$$\phi_q=\delta_{q,0}\sqrt{\beta L^d}\, r_0\, e^{i\theta}  \eqno (4)$$
where $\theta\in[0,2\pi]$ is an arbitrary phase and $\Delta^2=\lambda r_0^2$ is a 
solution of the BCS equation 
$${\ts {\lambda\over \beta L^d}}\sum_{k\in M_\nu} 
   \ts{1\over k_0^2+e_\k^2+\lambda r_0^2}=1   \eqno (5)$$
Equation (4) describes the BCS mean field configuration. 
\par
In Theorem 2  we expand $V$  up to second order in 
 $\xi=(\xi_{k-p})_{k,p}$ where 
$$\eqalignno{ \xi_{q}&=\phi_{q}-\delta_{q,0}\sqrt{\beta L^d}\,r_0\,
    e^{i\theta_0}  
  =\cases{ (\rho_0-\sqrt{\beta L^d}\,r_0)e^{i\theta_0} &for $q=0$ \cr 
   \rho_{q} e^{i\theta_{q}} &for $q\ne 0$. \cr}
  &(6)  \cr}$$
We remark that a priori there is no need to introduce a small external 
 field to fix the phase $\theta_0$ and then to expand with respect to 
 radial and tangential components as it is 
usually done (for example [2,3]). 
The expansion around $\xi$ gives 
$$\eqalignno{ V(\{\phi_q\})&=V_{\rm min}+2\beta_0 \,
  (\rho_0-\sqrt{\beta L^d}\,r_0)^2 +
  \sum_{q\ne 0} (\alpha_q+i\gamma_q)\, \rho_q^2 \cr
 &\phantom{m} + {\ts{1\over2}}\sum_{q\ne 0} 
   \beta_q \,|e^{-i\theta_0} \phi_q
   +e^{i\theta_0} \bar\phi_{-q}|^2 
  +O(\xi^3)   & (7) \cr} $$
where $V_{\rm min}\sim \beta L^d\Delta\log(1/\Delta)$ 
  is identical the global minimum of the BCS 
effective  potential 
$$V_{\rm BCS}(\rho)=\beta L^d\left(\rho^2-\ts {1\over\beta L^d}\sum_{k}
 \log\left[ 1+ {\lambda \rho^2\over k_0^2+e_\k^2}\right]
   \right)  \eqno (8)$$
and 
the coefficients $\alpha_q, \;\beta_q$ are real and positive 
  and $\gamma_q$ is real. They are  
given in Theorem 2 below. In particular, $\alpha_q\sim q^2$ for 
small $q$. 
\par
The case where  a small external U(1) symmetry breaking  field is
added to the fermionic action is discussed in section III. 
\par
The error term in (7) is not uniform in the coupling $\lambda$. 
Since $\beta_q\to 0$ and $\alpha_q+i\gamma_q\to 1$ 
   for $\lambda\to 0$ (see (54)), 
equation (7) becomes in the limit $\lambda\to 0$ 
$$ \sum_q|\phi_q|^2=\sum_{q\ne 0}|\phi_q|^2+O(\xi^3)
   \eqno (9)$$ 
or $|\phi_0|^2=O(\xi^3)$. The reason for this can already be seen 
if one expands $V_{\rm BCS}$ above around its minimum $r_0$. One 
obtains, abbreviating $E_k^2:=k_0^2+e_\k^2+\Delta^2$,   
$$\eqalignno{ {\ts {1\over \beta L^d}}\bigl(V_{\rm BCS}&(\rho)
  -V_{\rm min}\bigr) =\rho^2-r_0^2-\ts {1\over\beta L^d}\sum_k 
  \log\left[ 1+{\lambda(\rho^2-r_0^2)\over k_0^2+e_\k^2+\Delta^2}
  \right]   \cr
 &=\left\{1-\ts {\lambda\over \beta L^d}\sum_k{\ts {1\over 
   E_k^2}} \right\} (\rho^2-r_0^2)+ 
  \ts {1\over2} {1\over \beta L^d}\sum_k \ts  {\lambda^2\over 
   E_k^4}(\rho^2-r_0^2)^2 +O\bigl(
    (\rho-r_0)^3\bigr) \cr
 &=\ts {1\over2} {1\over \beta L^d}\sum_k \ts  {\lambda^2\over 
    E_k^4}\left[ 4r_0^2(\rho-r_0)^2+ 
  4r_0(\rho-r_0)^3+(\rho-r_0)^4\right] 
   +O\bigl( (\rho-r_0)^3\bigr)\;\;&(10)\cr}$$
where in the last line the BCS equation (5) has been used. 
Since ${1\over \beta L^d}\sum_k{1\over E_k^4}
  \sim{1\over\Delta^2} ={1\over\lambda r_0^2}$, the quadratic term 
 on the right hand side of (10) goes to 0, whereas the 
third and fourth 
order terms diverge. 
\bigskip
Thus it is not clear to what extent one may draw 
 conclusions from the quadratic approximation. Nevertheless, we
write down the results for the partition function and the
expectation value of the energy obtained by using (7). 
 To this end 
we briefly recall the relation between the Hamiltonian and the 
functional integral representation of the model. One may look in [5] 
to see more detailed computations. 
\bigskip
The Hamiltonian for the many-electron system with delta
interaction in finite volume at some small but 
 positive temperature $T={1\over \beta}>0$ described in the grand canonical 
 ensemble is given by  
$$H=H_0-\lambda H_{\rm int}={\ts
{1\over L^d}}\sum_{\k\sigma}(\vep_\k-\mu)a_{\k\sigma}^+ 
 a_{\k\sigma}-{\ts {\lambda\over L^{3d}}} \sum_{\k\p\q} a_{\k\up}^+ 
 a_{\q-\k\down}^+ a_{\q-\p\down} a_{\p\up} \eqno (11)$$
We are interested in the grand canonical partition function $Tr\,e^{-\beta H}$, 
 in the correlation function 
$$\tilde \Lambda(\q)={\ts {1\over L^{3d}}}\sum_{\k,\p}Tr[ a_{\k\up}^+ 
 a_{\q-\k\down}^+ a_{\q-\p\down} a_{\p\up} e^{-\beta H}]
     \bigr/ Tr\, e^{-\beta H}   \eqno (12)$$ 
  and in the expectation value 
$E_{\rm int}=\la H_{\rm int}\ra=\sum_\q \tilde \Lambda(\q)$. 
\medskip
In terms of Grassmann integrals, the perturbation series for the normalized 
partition function $Z=Tr e^{-\beta(H_0-\lambda H_{\rm int})}/Tr e^{-\beta H_0}$ 
is given by
$$Z=\int e^{ {\lambda\over(\beta L^d)^3}\sum_{k,p,q}\bar\psi_{k\up}
    \bar\psi_{q-k\down}\psi_{q-p\down}\psi_{p\up} } d\mu_C\eqno (13)$$
and $\tilde \Lambda(\q)$ is given by 
$$\tilde \Lambda(\q)={\ts {1\over \beta}}\sum_{q_0\in {2\pi\over\beta}\Bbb Z} 
  \Lambda(q_0,\q) \eqno (14)$$
where 
$$\eqalignno{ \Lambda(q_0,\q)&
   ={\ts {1\over (\beta L^d)^3}}\sum_{k,p}\la\bar\psi_{k\up}
    \bar\psi_{q-k\down}\psi_{q-p\down}\psi_{p\up}\ra &(15) \cr
  &={\ts {1\over (\beta L^d)^3}}\sum_{k,p} {\ts {1\over Z}}
 \int \bar\psi_{k\up}
    \bar\psi_{q-k\down}\psi_{q-p\down}\psi_{p\up}\>
  e^{ {\lambda\over(\beta L^d)^3}\sum_{k,p,q}\bar\psi_{k\up}
    \bar\psi_{q-k\down}\psi_{q-p\down}\psi_{p\up} } d\mu_C \cr 
 &={\ts {d\over d\lambda_q }}_{|_{\lambda_q=\lambda}}\log \int 
    e^{ {1\over(\beta L^d)^3}\sum_{k,p,q}\lambda_q \bar\psi_{k\up}
    \bar\psi_{q-k\down}\psi_{q-p\down}\psi_{p\up} } d\mu_C&(16) \cr}$$
By making a Hubbard Stratonovich  transformation, that is, by applying the 
formula ($\phi_q=u_q+iv_q$, $\bar\phi_q=u_q-iv_q$, $d\phi_qd\bar\phi_q:=
  du_qdv_q$) 
$$  e^{\sum_q a_q b_q}=\int e^{\sum_q a_q\phi_q+b_q\bar\phi_q} 
  e^{-\sum_q|\phi_q|^2}\pro_q{\ts {d\phi_qd\bar\phi_q\over \pi}}$$
with $a_q=ig_q(\beta L^d)^{-{3\over2}} 
      \sum_k\bar\psi_{k\up}\bar\psi_{q-k\down}$ 
 and $b_q=ig_q(\beta L^d)^{-{3\over2}} 
     \sum_p \psi_{p\up}\psi_{q-p\down}$, $g_q=\sqrt\lambda_q$, 
  the exponent becomes quadratic in
the Grassmann variables and the fermionic integral can be performed. 
Since the Grassmann variables in the exponent can be arranged,   
 $(\bar\psi_\up,\psi_\down)$ and $(\psi_\up,\bar\psi_\down)$, such
that a given $\psi$ appears only in one but not in both factors, 
 the Pfaffian comming from the fermionic integration reduces 
 to  a determinant. The 
 result is 
$$Z(\{\lambda_q\})=\int e^{-V(\{\phi_q\})} \pro_q 
         {\ts {d\phi_qd\bar\phi_q\over \pi}} \eqno (17)$$
where, if the $\lambda_q$'s are not all equal, instead of (1) $V$ is given by 
$$\eqalignno{ V(\{\phi_q\})&=\sum_q |\phi_q|^2  \ts
  -\log{ \det\left[\matrix{ Id +{1\over\beta L^d}
 C (g\phi)^*  \bar C(g\phi)}\right]  } & (18) \cr}$$
where $(g\phi)$ stands for the matrix $(g_{k-p}\phi_{k-p})_{k,p}$. To perform 
the $\lambda_q$-derivative, one may change variables to obtain 
$$\eqalignno{ \Lambda(q)&={\ts {1\over Z}{d\over d\lambda_q}} \int 
  e^{-\sum_q {|\phi_q|^2\over\lambda_q}}\ts
    \det\left[\matrix{  Id+{1\over\beta L^d} \bar C\phi C \phi^*}\right]
   \pro_q {\ts {d\phi_qd\bar\phi_q\over\lambda_q \pi}}  \cr
 &={\ts {1\over Z}}  \int\left(  {\ts {|\phi_q|^2\over \lambda_q^2}
   -{1\over\lambda_q} }\right) 
  e^{-\sum_q {|\phi_q|^2\over\lambda}} \det\left[\matrix{  
     Id+{1\over\beta L^d} \bar C\phi C \phi^*}\right]
 \pro_q {\ts {d\phi_qd\bar\phi_q\over\lambda\pi}}  \cr
 &={\ts {1\over Z}}  \int {\ts {|\phi_q|^2-1\over \lambda}} 
  e^{-\sum_q {|\phi_q|^2}} \det\left[\matrix{
    Id+{\lambda\over\beta L^d} \bar C\phi C \phi^*    }\right] 
  \pro_q {\ts {d\phi_qd\bar\phi_q\over\pi}} \cr
  &= {\ts {1\over Z}}  \int {\ts {|\phi_q|^2-1\over \lambda}} e^{-V(\{\phi_q\})} 
   \pro_q {\ts {d\phi_qd\bar\phi_q\over\pi}}&(19)  \cr}$$
For the ideal Fermi gas $\lambda=0$  the above formula gives ${0\over0}$. 
  In that case one may use (15) to obtain 
$$\eqalignno{ \Lambda(q)& ={\ts{1\over (\beta L^d)^3}}\sum_{k,p} 
    \la\bar \psi_{k\up}\psi_{p\up}\ra  
    \la \bar\psi_{q-k\down}\psi_{q-p\down}\ra 
={\ts{1\over (\beta L^d)^3}}\sum_{k,p} \beta L^d \delta_{k,p} C_k
    \>\beta L^d \delta_{q-k,q-p} C_{q-k} \cr
 &={\ts{1\over \beta L^d}}\sum_k C_k C_{q-k} &(20) \cr}$$ 
which is the particle particle bubble. 
\bigskip
If one replaces $V$ with the second order approximation 
$$V_2(\{\phi_q\}):=V_{\rm min}+2\beta_0 \,
  (\rho_0-\sqrt{\kappa}\,r_0)^2 +
  \sum_{q\ne 0}( \alpha_q+i\gamma_q) \rho_q^2 
  + {\ts{1\over2}}\sum_{q\ne 0} \beta_q \,|e^{-i\theta_0} \phi_q
   +e^{i\theta_0} \bar\phi_{-q}|^2 \eqno (21)$$
the integrals in (17,19) become Gaussian  and can be performed.
  The results 
are (the index ``2" in the following means that $V_2$ instead of $V$ has been 
 used) 
$$Z_2=e^{-V_{\rm min}} \,
  \int_0^\infty e^{-2\beta_0^2(\rho_0-\sqrt\kappa 
  r_0)^2 } 2\rho_0d\rho_0 
  \pro_{q\atop q_0>0} 
  \ts {1\over \alpha_q^2+\gamma_g^2+2\alpha_q\beta_q} \>,
    \eqno (22)$$
$$\Lambda_2(q)=\ts {1\over\lambda} \left( {\alpha_q+i\gamma_q+\beta_q \over 
    \alpha_q^2+\gamma_q^2+2\alpha_q\beta_q} -1 
    \right)\eqno (23) $$
and, since $\gamma_{-q}=-\gamma_q$, 
$$\eqalignno{ \varepsilon_{\rm int,2}&:={\ts {1\over L^d}}
   \la H_{\rm int}\ra_2
   ={\ts {1\over \beta L^d }} 
  \sum_q \Lambda_2(q) 
  ={\ts {1\over \lambda}{1\over \beta L^d}} \sum_q {\ts 
   \left( {\alpha_q+\beta_q \over 
    \alpha_q^2+\gamma_q^2+2\alpha_q\beta_q} -1 \right)}
   &(24) \cr} $$
In particular, since $\alpha_q\le const_\lambda\,  q^2$ and 
 $|\gamma_q|\le const_\lambda\,  |q|$  for small $q$, 
$$\varepsilon_{\rm int,2}\ge  {\ts {1\over \lambda}
   {1\over\beta L^d}} \sum_q \ts 
   \left(  {const_\lambda \over q^2}-1\right) \eqno (25) $$ 
which, for $L\to\infty$, 
  is infrared singular for $d=1$ and, since $q_0=0$ is an allowed 
 value for positive temperature, 
 also logarithmically divergent for small $q$ in 
 $d=2$. A similar observation is made in [3].  
%
%
%
%
\bigskip
\bigskip
\noindent{\gross II. Proof of Theorems} 
\bigskip
\noindent{\bf Theorem 1:} {\it Let {\rm Re}$V$ be the real part of the effective
potential for the many electron system with attractive delta 
 interaction given by (3). Let $e_\k=\vep_\k-\mu$ satisfy 
 $e_\k=e_{-\k}$ and: $\forall\k\;e_\k=e_{\k+\q}\Rightarrow \q=0$. 
  Then all global minima of 
 Re$V$ are given by (4),
$$\phi_q=\delta_{q,0}\sqrt{\beta L^d} \, r_0\, 
     e^{i\theta},\1cm \theta\in[0,2\pi]\;\;\;{\rm 
 arbitrary} \eqno (4)$$
where $r_0$ is a solution of the BCS equation (5) 
or, equivalently, the global minimum of the function (8)
$$\eqalignno{ V_{\rm BCS}(\rho)&=V(\{\delta_{q,0}\sqrt{\beta
   L^d}\rho\,e^{i\theta} \}) \cr
 &= \beta L^d\rho^2-\sum_k 
   {\ts\log\left[ 1+{\lambda \rho^2\over k_0^2+e_\k^2}\right]} 
  =\beta L^d \rho^2-
  \sum_\k {\ts\log\left[{ \cosh({\beta\over2}
    \sqrt{ e_\k^2+\lambda\rho^2})\over 
    \cosh{\beta\over2}e_\k }\right]} & (8)\cr}$$
More specifically, there is the bound 
$$\eqalignno{  {\rm Re}V(\{\phi_q\})\ge \ts  V_{\rm BCS}(
    \|\phi\|)& -\min_{k\in M_\nu}\ts  \Biggl\{ 
   \log\left[{\ds \pro_{q\ne 0} }
  \biggl(1-{|{\lambda\over \beta L^d}\sum_{p}\phi_p 
  \bar \phi_{p+q}|^2\over (|a_k|^2+\lambda\|\phi\|^2)
   (|a_{k-q}|^2+\lambda\|\phi\|^2) } \biggr)^{1\over2}
    \right] &(26) \cr
  &\phantom{mmmm}+\ts\log\left[{\ds \pro_{q\ne 0} }
  \biggl(1-{ {\lambda\over \beta L^d}|\phi_q|^2|a_{k}-a_{k-q}|^2
  \over (|a_k|^2+\lambda\|\phi\|^2)
   (|a_{k-q}|^2+\lambda\|\phi\|^2) } \biggr)^{1\over2} \right]\Biggr\}
   \cr}  $$
where $\|\phi\|^2:={1\over\beta L^d}\sum_q|\phi_q|^2$ and $|a_k|^2:= 
  k_0^2+e_\k^2$. In particular,
$${\rm Re}V(\{\phi_q\})\ge  V_{\rm BCS}( \|\phi\|)\eqno (27)$$
since the products in (26) are less or equal 1.    }
\bigskip
\noindent{\bf Proof:} Suppose first that (26) holds. 
 For each $q$, the round brackets 
in (26) are between 0 and 1 which means that $-\log(\pro_q\cdots)$ is positive. 
Thus 
$${\rm Re}V(\{\phi_q\})\ge 
  \ts  V_{\rm BCS}( \|\phi\|)\ge V_{\rm BCS}(r_0)\eqno (28)$$ 
which proves that $\phi_q=\delta_{q,0}\sqrt{\beta  L^d}r_0\,e^{i\theta} $ are indeed 
global minima of Re$V$. On the other hand, if a configuration $\{\phi_q\}$ 
is a global minimum, then the logarithms in (26) must be zero for all 
 $k\in M_\nu$ which in particular 
 means that for all $q\ne 0$ 
$$\sum_{p}\phi_p  \bar \phi_{p+q}=0 \eqno(29)$$
and for all $k\in M_\nu$ and $q\ne 0$
$$  |\phi_q|^2|a_{k}-a_{k-q}|^2
  =|\phi_q|^2[q_0^2+(e_{\k}-e_{\k-\q})^2]=0\eqno(30)$$
which implies $\phi_q=0$ for all $q\ne 0$. 
  It remains to prove (26). 
\par
To this end, we write (recall that $C_k={1\over a_k}={1\over ik_0-e_\k}$)
$$\det\left[\matrix{ Id & {ig\over 
  \sqrt{\beta L^d} }\, 
 \bar C \phi^* \cr {ig\over\sqrt{\beta L^d} }\, C\phi  & Id \cr}\right]=\det\left[ 
  \matrix{&|& &|& \cr &\vec b_k& &\vec b'_{k'}& \cr &|& &| & \cr}\right]
   \eqno (31)$$
where ($k,k'$ fixed, $p$ labels the vector components)
$$\vec b_k=\pmatrix{ \cr \delta_{k,p}\cr \cr
    {\ts {ig\over\sqrt{\beta L^d}}}
  {\phi_{k-p}\over  a_k} \cr \cr}\1cm 
  \vec b'_{k'}= 
  \pmatrix{ \cr {\ts {ig\over\sqrt{\beta L^d}}}{\bar\phi_{p-k'}\over 
   \bar a_{k'}}\cr \cr \delta_{k',p} \cr \cr}  \eqno (32) $$
If $|\vec b_k|$ denotes the euclidean norm of $\vec b_k$, then we have 
$$\ts |\vec b_k|^2=1+{\lambda\over\beta L^d}\sum_p{|\phi_{k-p}|^2\over 
  |a_k|^2}=1+ \lambda{\|\phi\|^2\over 
   |a_k|^2}=|\vec b'_{k}|^2 \eqno (33)$$
Therefore one obtains, if $\vec e_k={\vec b_k\over |\vec b_k|}$, 
 ${\vec e}^{\,\prime}_k={\vec b'_k\over |\vec b'_k|}$
$$\det\left[ 
  \matrix{&|& &|& \cr &\vec b_k& &\vec b'_{k'}& \cr &|& &| & \cr}\right]=
  \pro_k\left\{ 1+ \lambda{\|\phi\|^2\over  |a_k|^2} \right\}
  \det\left[ 
  \matrix{&|& &|& \cr &\vec e_k& &{\vec e}^{\,\prime}_{k'}
   & \cr &|& &| & \cr}\right]  \eqno (34)$$
From this the inequality (27) already follows since the 
 determinant on the right hand side of (34) is  less or equal 1. 
To obtain (26), 
   we choose a fixed but arbitrary momentum $t\in M_\nu$ and 
 orthogonalize all vectors $\vec e_k$, ${\vec e}^{\,\prime}_{k'}$ in the determinant with 
 respect to $e_t$. That is, we write 
$$ \det\left[ 
  \matrix{|&|& &|& \cr \vec e_t &\vec e_k& &{\vec e}^{\,\prime}_{k'}
  & \cr | &|& &| & \cr}\right]= 
  \det\left[ \matrix{|&|& &|& \cr \vec e_t& 
  \vec e_k-(\vec e_k,\vec e_t)\vec e_t& 
  &{\vec e}^{\,\prime}_{k'}-({\vec e}^{\,\prime}_{k'},\vec e_t)
  \vec e_t
    & \cr |&|& &| & \cr}\right]  \eqno(35)$$
Finally we apply Hadamard's inequality, $ |\det F|\le \pro_{j=1}^n |\vec f_j|
  =\Bigl\{ \pro_{j=1}^n \sum_{i=1}^n |f_{ij}|^2 
  \Bigr\}^{1\over2} $ if $F=(f_{ij})_{1\le i,j\le n}$ is a complex matrix, to the 
determinant on the right hand side of (35). Since 
$$|\vec e_k-(\vec e_k,\vec e_t)\vec e_t|^2=1-|(\vec e_k,\vec e_t)|^2$$
one obtains 
$$\left|\det\left[ \matrix{&|& &|& \cr &\vec e_k-(\vec e_k,\vec e_t)\vec e_t& 
  &{\vec e}^{\,\prime}_{k'}-({\vec e}^{\,\prime}_{k'},\vec e_t)\vec e_t
   & \cr &|& &| & \cr}\right]\right| $$
$$  \le |\vec e_t|\pro_{k\in M_\nu\atop k\ne t} (1-|(\vec e_k,\vec
    e_t)|^2)^{1\over2} 
  \pro_{k\in M_\nu}
  (1-|({\vec e}^{\,\prime}_k,\vec e_t)|^2)^{1\over2} \eqno(36)$$ 
or with (31) and (34) 
$$\eqalignno{ {\rm Re}V(\{\phi_q\})&=\sum_q|\phi_q|^2
   -\log \left|\det\left[\matrix{ Id & {ig\over 
  \sqrt{\beta L^d} }\, 
  \phi^*\bar C \cr {ig\over\sqrt{\beta L^d} }\, \phi C & Id \cr}\right]\right|\cr
 &=\beta L^d\|\phi\|^2 -\sum_k\log
   \left\{ 1+ \lambda{\|\phi\|^2\over  |a_k|^2} \right\}
  -\log\left|
   \det\left[ \matrix{&|& &|& \cr &\vec e_k& &{\vec e}^{\,\prime}_{k'}& 
     \cr &|& &| & \cr}\right]  \right|\cr 
&=V_{\rm BCS}(\|\phi\|) -\log\left|
   \det\left[ \matrix{&|& &|& \cr &\vec e_k& &{\vec e}^{\,\prime}_{k'}& 
     \cr &|& &| & \cr}\right]  \right|\cr 
 &\ge V_{\rm BCS}(\|\phi\|)-\log \Bigl\{ 
   \pro_{k\in M_\nu\atop k\ne t} (1-|(\vec e_k,\vec e_t)|^2)^{1\over2} 
  \pro_{k\in M_\nu}
 (1-|({\vec e}^{\,\prime}_k,\vec e_t)|^2)^{1\over2}\Bigr\}&(37) 
     \cr}$$ 
Finally one has 
$$(\vec b_k,\vec b_t)=\sum_p {\ts {ig\over\sqrt{\beta L^d}}} 
   {\phi_{k-p}\over  a_k}\> \overline{ {\ts {ig\over\sqrt{\beta L^d}}} 
      {\phi_{t-p}\over  a_t} }$$
which gives 
$$|(\vec e_k,\vec e_t)|^2=\ts {\left| {\lambda\over\beta L^d}\sum_p\phi_{k-p}
     \bar\phi_{t-p}\right|^2\over (|a_k|^2+\lambda\|\phi\|^2)\phantom{^{1\over2}}
   \!\!\!(|a_t|^2+\lambda\|\phi\|^2) }
 =\ts {\left| {\lambda\over\beta L^d}\sum_p \phi_{p}
     \bar\phi_{t-k+p}\right|^2\over (|a_k|^2+\lambda\|\phi\|^2)\phantom{^{1\over2}}
   \!\!\!(|a_t|^2+\lambda\|\phi\|^2) } \eqno (38a)$$ 
and 
$$\eqalignno{ (\vec b'_k,\vec b_t)&
    ={\ts {ig\over\sqrt{\beta L^d}}}{\bar\phi_{t-k}\over \bar a_{k}}+
  \overline{  {\ts {ig\over\sqrt{\beta L^d}}}  {\phi_{t-k}\over  a_t} } 
  = {\ts {ig\over\sqrt{\beta L^d}}}\bar\phi_{t-k}\ts \left({1\over \bar a_{k}}-
     {1\over \bar a_{t}} \right) \cr}$$
which gives
$$|({\vec e}^{\,\prime}_k,\vec e_t)|^2=\ts { {\lambda\over\beta L^d}|\phi_{t-k}|^2
   |a_t-a_k|^2 \over (|a_k|^2+\lambda\|\phi\|^2)\phantom{^{1\over2}}
   \!\!\!(|a_t|^2+\lambda\|\phi\|^2) } \eqno (38b)$$
Substituting (38a,b) in (37) gives, substituting $k\to q=t-k$  
$$\eqalignno{ {\rm Re}V(\{\phi_q\})&\ge
   V_{\rm BCS}(\|\phi\|)-\log \Bigl\{ 
   \pro_{k\in M_\nu\atop k\ne t} (1-|(\vec e_k,\vec e_t)|^2)^{1\over2} 
  \pro_{k\in M_\nu}
 (1-|({\vec e}^{\,\prime}_k,\vec e_t)|^2)^{1\over2}\Bigr\}\cr  
 &\buildrel k=t-q\over   =V_{\rm BCS}(\|\phi\|)-\log \Bigl\{
   \pro_{q\ne 0}  (1-|(\vec e_{t-q},\vec e_t)|^2)^{1\over2} 
 \pro_{q\ne 0}(1-|({\vec e}^{\,\prime}_{t-q},\vec e_t)|^2)^{1\over2}\Bigr\}
  &(39)\cr}$$
Since $t$ was arbitrary, we can take the maximum of the right hand
side of (39) with respect to $t$ which  
 proves Theorem 1 $\blacksquare$
\bigskip
%
%
%
%
\bigskip
\noindent{\bf Theorem 2:} {\it  Let $V$ be the effective potential  (1), let 
 $\kappa=\beta L^d$  and let 
$\xi=(\xi_{k-p})_{k,p}$ be the matrix with entries  
$$\eqalignno{ \xi_{q}&=\phi_{q}-\delta_{q,0}\sqrt{\kappa}\,r_0\,
    e^{i\theta_0} =\cases{ (\rho_0-\sqrt{\kappa}\,r_0)e^{i\theta_0} &for $q=0$ \cr 
   \rho_{q} e^{i\theta_{q}} &for $q\ne 0$. \cr} &(40) \cr}$$
Then
$$ V(\{\phi_q\})=V_{\rm min}+2\beta_0 \,
  (\rho_0-\sqrt{\kappa}\,r_0)^2 +
  \sum_{q\ne 0}( \alpha_q+i\gamma_q) \rho_q^2 
  + {\ts{1\over2}}\sum_{q\ne 0} \beta_q \,|e^{-i\theta_0} \phi_q
   +e^{i\theta_0} \bar\phi_{-q}|^2 
  +O(\xi^3)   \eqno(41)  $$
where, if $E_k^2=k_0^2+e_\k^2+\lambda r_0^2$, 
$$\alpha_q=
  {\ts {1\over2}}{\ts{\lambda\over\kappa}}\sum_k 
  {\ts { q_0^2+(e_\k-e_{\k-\q})^2 \over E_k^2 E_{k-q}^2}} >0\,,\;\;\;
\beta_q= {\ts{\lambda\over\kappa}}\sum_k \ts {\lambda r_0^2 
  \over E_k^2 E_{k-q}^2} >0\,,\;\;$$
$$\gamma_q
   =-{\ts {\lambda\over  \kappa }} \sum_k \ts 
  {k_0e_{\k-\q}-(k_0-q_0)e_\k\over E_k^2 E_{k-q}^2} 
   \in \Bbb R   \eqno (42)$$
and 
$$\ts V_{\rm min}=\kappa\left( r_0^2-
 {1\over\kappa} \sum_\k {\ts\log\left[{ \cosh({\beta\over2}
    \sqrt{ e_\k^2+\lambda r_0^2})\over 
    \cosh{\beta\over2}e_\k }\right]}\right).  \eqno (43)$$  } 
\bigskip
\noindent{\bf Proof:} We abbreviate $\kappa=\beta L^d$ and write 
$$V(\{\phi_q\})=\sum_q |\phi_q|^2-\log{ \det\left[\matrix{ A& {ig\over\wk} 
  \phi^* \cr {ig\over\wk} \phi & \bar A \cr}\right] \over 
  \det\left[\matrix{ A& 0 \cr 0 & \bar A \cr}\right] } \eqno
  (44)$$
where $A=C^{-1}=(\delta_{k,p}a_k)_{k,p\in M_\kappa}$ and
   $a_k:=1/C_k=
  ik_0-e_\k$. Then 
$$ \eqalignno{ V(\{\phi_q\})- V(\{\sqrt\kappa \, \delta_{q,0}\, r_0 e^{i\theta_0} \})
  &=\sum_q \rho_q^2 -\kappa r_0^2
   -\log{ \det\left[\matrix{ A& {ig\over\wk} 
  \phi^* \cr {ig\over\wk} \phi & \bar A \cr}\right] \over 
  \det\left[\matrix{ A& igr_0\, e^{-i\theta_0} \cr 
  igr_0\, e^{i\theta_0} & \bar A \cr}\right] }  &(45) \cr}$$
where $ igr_0\, e^{i\theta_0}\equiv  igr_0\, e^{i\theta_0}\,Id$ in the 
 determinant above. 
 Since 
$$\eqalignno{ \left[\matrix{ A& igr_0\, e^{-i\theta_0} \cr 
     igr_0\, e^{i\theta_0} & \bar A \cr}\right]^{-1}& =
\left[\matrix{ {\bar a_k\delta_{k,p}\over |a_k|^2+\lambda r_0^2} & 
 -{ igr_0\, e^{-i\theta_0}\delta_{k,p}\over |a_k|^2+\lambda r_0^2}  \cr 
   - { igr_0\, e^{i\theta_0} \delta_{k,p}\over |a_k|^2+\lambda r_0^2} & 
 {a_k \delta_{k,p}\over |a_k|^2+\lambda r_0^2}  \cr}\right]  \cr
  &\equiv 
{\ts {1\over|a|^2+\lambda r_0^2}}
 \left[\matrix{ \bar A& -igr_0\, e^{-i\theta_0} \cr 
     -igr_0\, e^{i\theta_0} & A \cr}\right]  &(46)  \cr}$$
and because of 
$$\eqalignno{ \left[\matrix{ A& {ig\over\wk} 
  \phi^* \cr {ig\over\wk} \phi & \bar A \cr}\right]&= 
 \left[\matrix{ A& igr_0\, e^{-i\theta_0}
   \cr  igr_0\, e^{i\theta_0}   & \bar A \cr}\right]\>+\> 
  \left[\matrix{ 0& {ig\over\wk} 
  \phi^*-igr_0\, e^{-i\theta_0} \cr {ig\over\wk} \phi -igr_0\, e^{i\theta_0}&
    0\cr}\right] \cr
 &= \left[\matrix{ A& ig \bar\gamma
   \cr  ig\gamma   & \bar A \cr}\right]\>+\> 
  \left[\matrix{ 0& ig\,\xi^* \cr  ig\,\xi &
    0\cr}\right] &(47)\cr}$$
where $\gamma=r_0\, e^{i\theta_0}$ and $\xi=(\xi_{k,p})$ 
  is given by (40), 
the quotient of determinants in (45) is given by 
$$\eqalignno{ \det&\left[ Id+  {\ts {1\over|a|^2+\lambda r_0^2}}
 \left(\matrix{ \bar A& -ig\bar\gamma  \cr 
     -ig\gamma  & A \cr}\right)  \left(\matrix{ 0& ig\,\xi^* \cr  ig\,\xi &
    0\cr}\right)\right] \cr
  &\phantom{mmmmmmm}= \det\left[ Id+ \left(\matrix{ \lambda
 {\bar\gamma \over|a|^2+\lambda r_0^2}\,\xi &
    ig {\bar A \over|a|^2+\lambda r_0^2}\,\xi^* \cr 
   ig {A \over|a|^2+\lambda r_0^2}\,\xi &
    \lambda {\gamma \over|a|^2+\lambda r_0^2}\,\xi^* \cr}\right)\right]
   & (48)\cr}$$
Since 
$$\eqalignno{ \log\det[Id+B]&=Tr\log[Id+B]
     =\sum_{n=1}^\infty {(-1)^{n+1}\over n} Tr B^n  \cr
 &=Tr\,B-{\ts{1\over2}}Tr\, B^2+{\ts{1\over3}} 
  Tr\, B^3-+\cdots &(49) \cr}$$
one obtains to second order in $\xi$:
$$\eqalignno{ \log&\det\left[ Id+ \left(\matrix{ \lambda
 {\bar\gamma \over|a|^2+\lambda r_0^2}\,\xi &
    ig {\bar A \over|a|^2+\lambda r_0^2}\,\xi^* \cr 
  ig {A \over|a|^2+\lambda r_0^2}\,\xi &
    \lambda {\gamma \over|a|^2+\lambda r_0^2}\,\xi^* \cr}
  \right)\right] \cr
 &=Tr\left(\matrix{ \lambda
 {\bar\gamma \over|a|^2+\lambda r_0^2}\,\xi &
    ig {\bar A \over|a|^2+\lambda r_0^2}\,\xi^* \cr 
   ig {A \over|a|^2+\lambda r_0^2}\,\xi &
    \lambda {\gamma \over|a|^2+\lambda r_0^2}\,\xi^* \cr}\right)\>-\>
  {\ts{1\over2}}Tr\biggl\{ \left(\matrix{ \lambda
 {\bar\gamma \over|a|^2+\lambda r_0^2}\,\xi &
    ig {\bar A \over|a|^2+\lambda r_0^2}\,\xi^* \cr 
   ig {A \over|a|^2+\lambda r_0^2}\,\xi &
    \lambda {\gamma \over|a|^2+\lambda r_0^2}\,\xi^* \cr}\right)^2 \biggr\} 
    +O(\xi^3) \cr
 &=\ts  Tr
 {\lambda\bar\gamma \over|a|^2+\lambda r_0^2}\,\xi 
  +Tr  {\lambda\gamma \over|a|^2+\lambda r_0^2}\,\xi^*
  -{\ts{1\over2}}\biggl\{ Tr
 {\lambda\bar\gamma \over|a|^2+\lambda r_0^2}\,\xi
 {\lambda\bar\gamma \over|a|^2+\lambda r_0^2}\,\xi
  +Tr  {ig\bar A \over|a|^2+\lambda r_0^2}\,\xi^*
   {ig A \over|a|^2+\lambda r_0^2}\,\xi   \cr
 &\phantom{=}\ts +
  Tr  {\lambda\gamma \over|a|^2+\lambda r_0^2}\,\xi^*
    {\lambda\gamma \over|a|^2+\lambda r_0^2}\,\xi^*
  +Tr  { ig A \over|a|^2+\lambda r_0^2}\,\xi
   {ig \bar A \over|a|^2+\lambda r_0^2}\,\xi^*
    \biggr\}+O(\xi^3)&(50) \cr}$$
One has 
$$\ts \left({\lambda\bar\gamma \over|a|^2
  +\lambda r_0^2}\,\xi\right)_{k,p}
 ={\lambda \bar\gamma \over|a_k|^2
  +\lambda r_0^2}\,\xi_{k,p}\>, \;\;\;\; 
  \left({\lambda\gamma \over|a|^2+\lambda r_0^2}\,\xi^*\right)_{k,p} 
  ={\lambda\gamma \over|a_k|^2+\lambda r_0^2}\,\bar\xi_{p,k}$$
$$\eqalignno{ \ts \left({\lambda\bar\gamma \over|a|^2
  +\lambda r_0^2}\,\xi{\lambda\bar\gamma \over|a|^2
  +\lambda r_0^2}\,\xi\right)_{k,k}&=\sum_p \ts 
  {\lambda \bar\gamma \over|a_k|^2+\lambda r_0^2}\,\xi_{k,p}
  {\lambda \bar\gamma \over|a_p|^2+\lambda r_0^2}\,\xi_{p,k} \cr
  &= {\ts 
  \left({\lambda \bar\gamma \,\xi_{k,k} \over|a_k|^2+\lambda
   r_0^2}\right)^2  }
  +\sum_{p\atop p\ne k} \ts 
  {\lambda \bar\gamma \over|a_k|^2+\lambda r_0^2}
  {\lambda \bar\gamma \over|a_p|^2+\lambda r_0^2}\,\xi_{k,p}
  \,\xi_{p,k} \cr
 &= {\ts {1\over\kappa} 
  \left({\lambda r_0 (\rho_0-\sqrt\kappa r_0) \over|a_k|^2+\lambda
   r_0^2}\right)^2  }
  +{\ts {1\over\kappa}}\sum_{q\ne 0} \ts 
  {\lambda \bar\gamma \over|a_k|^2+\lambda r_0^2}
  {\lambda \bar\gamma \over|a_{k-q}|^2+\lambda r_0^2}\,\phi_q
  \,\phi_{-q} \cr}$$
$$\eqalignno{ \ts \left({\lambda\gamma \over|a|^2
  +\lambda r_0^2}\,\xi^*{\lambda\gamma \over|a|^2
  +\lambda r_0^2}\,\xi^*\right)_{k,k}&=\sum_p \ts 
  {\lambda \gamma \over|a_k|^2+\lambda r_0^2}\,\bar\xi_{p,k}
  {\lambda \gamma \over|a_p|^2+\lambda r_0^2}\,\bar\xi_{k,p} \cr
  &= {\ts 
  \left({\lambda \gamma \,\bar\xi_{k,k} \over|a_k|^2+\lambda
   r_0^2}\right)^2  }
  +\sum_{p\atop p\ne k} \ts 
  {\lambda \gamma \over|a_k|^2+\lambda r_0^2}
  {\lambda \gamma \over|a_p|^2+\lambda r_0^2}\,\bar\xi_{k,p}
  \,\bar\xi_{p,k} \cr
 &= {\ts {1\over\kappa} 
  \left({\lambda r_0 (\rho_0-\sqrt\kappa r_0) \over|a_k|^2+\lambda
   r_0^2}\right)^2  }
  +{\ts {1\over\kappa}}\sum_{q\ne 0} \ts 
  {\lambda \gamma \over|a_k|^2+\lambda r_0^2}
  {\lambda \gamma \over|a_{k-q}|^2+\lambda r_0^2}\,\bar\phi_q
  \,\bar\phi_{-q} \cr}$$
and 
$$\ts \left({ ig A \over|a|^2+\lambda r_0^2}\,\xi\right)_{k,p}=
  { ig a_k \over|a_k|^2+\lambda r_0^2}\,\xi_{k,p}\>,\;\;\;\; 
  \ts \left({ ig \bar A \over|a|^2+\lambda r_0^2}\,\xi^*\right)_{k,p}=
  { ig \bar a_k \over|a_k|^2+\lambda r_0^2}\,\bar\xi_{p,k}$$
$$\eqalignno{\ts \left({ ig A \over|a|^2+\lambda r_0^2}\,\xi
   { ig \bar A \over|a|^2+\lambda r_0^2}\,\xi^* \right)_{k,k}&=
  \sum_p \ts { ig a_k \over|a_k|^2+\lambda r_0^2}\,\xi_{k,p}
  { ig \bar a_p \over|a_p|^2+\lambda r_0^2}\,\bar\xi_{k,p}\cr
  &= -\lambda{\ts {  |a_k|^2  
     \over(|a_k|^2+\lambda r_0^2)^2} }|\xi_{k,k}|^2
 -\lambda \sum_{p\atop p\ne k}
   \ts {  a_k \over|a_k|^2+\lambda r_0^2}\,\xi_{k,p}
  {  \bar a_p \over|a_p|^2+\lambda r_0^2}\,\bar\xi_{k,p}\cr
  &={\ts -{\lambda\over\kappa} {  |a_k|^2  
     \over(|a_k|^2+\lambda r_0^2)^2} }(\rho_0-\sqrt\kappa r_0)^2
 -{\ts {\lambda\over\kappa}} \sum_{q\ne 0}
  \ts {  a_k \over|a_k|^2+\lambda r_0^2}\,
  {  \bar a_{k-q} \over|a_{k-q}|^2+\lambda r_0^2}\, \rho_q^2\cr}$$
$$\eqalignno{\ts \left({ ig \bar A \over|a|^2+\lambda r_0^2}\,\xi^*
   { ig  A \over|a|^2+\lambda r_0^2}\,\xi \right)_{k,k}&=
  \sum_p \ts { ig \bar a_k \over|a_k|^2+\lambda r_0^2}\,\bar\xi_{p,k}
  { ig  a_p \over|a_p|^2+\lambda r_0^2}\,\xi_{p,k}\cr
  &= -\lambda{\ts {  |a_k|^2  
     \over(|a_k|^2+\lambda r_0^2)^2} }|\xi_{k,k}|^2
 -\lambda \sum_{p\atop p\ne k}
   \ts {  \bar a_k \over|a_k|^2+\lambda r_0^2}\,\bar\xi_{p,k}
  {   a_p \over|a_p|^2+\lambda r_0^2}\,\xi_{p,k}\cr
  &={\ts -{\lambda\over\kappa} {  |a_k|^2  
     \over(|a_k|^2+\lambda r_0^2)^2} }(\rho_0-\sqrt\kappa r_0)^2
 -{\ts {\lambda\over\kappa}} \sum_{q\ne 0}
   \ts {  \bar a_k \over|a_k|^2+\lambda r_0^2}\,
  {   a_{k+q} \over|a_{k+q}|^2+\lambda r_0^2}\, \rho_q^2\cr}$$
Therefore (50) becomes 
$$\eqalignno{ \log&\det\left[ Id+ \left(\matrix{ \lambda
 {\bar\gamma \over|a|^2+\lambda r_0^2}\,\xi &
    ig {\bar A \over|a|^2+\lambda r_0^2}\,\xi^* \cr 
   ig {A \over|a|^2+\lambda r_0^2}\,\xi &
    \lambda {\gamma \over|a|^2+\lambda r_0^2}\,\xi^* \cr}\right)\right] \cr
 &=\ts  Tr
 {\lambda\bar\gamma \over|a|^2+\lambda r_0^2}\,\xi 
  +Tr  {\lambda\gamma \over|a|^2+\lambda r_0^2}\,\xi^*
  -{\ts{1\over2}}\biggl\{ Tr
 {\lambda\bar\gamma \over|a|^2+\lambda r_0^2}\,\xi
 {\lambda\bar\gamma \over|a|^2+\lambda r_0^2}\,\xi
  +Tr  {ig\bar A \over|a|^2+\lambda r_0^2}\,\xi^*
   {ig A \over|a|^2+\lambda r_0^2}\,\xi   \cr
 &\phantom{=}\ts +
  Tr  {\lambda\gamma \over|a|^2+\lambda r_0^2}\,\xi^*
    {\lambda\gamma \over|a|^2+\lambda r_0^2}\,\xi^*
  +Tr  { ig A \over|a|^2+\lambda r_0^2}\,\xi
   {ig \bar A \over|a|^2+\lambda r_0^2}\,\xi^*
    \biggr\}+O(\xi^3) \cr
 &=2{\ts{\lambda\over\kappa}}\sum_k{\ts {\sqrt\kappa\,r_0(\rho_0-
  \sqrt\kappa\,r_0)\over |a_k|^2+\lambda r_0^2}} \cr
 &\phantom{=}  -{\ts {1\over2}} \biggl\{
    {\ts {1\over\kappa}} \sum_k {\ts
  \left({\lambda r_0 (\rho_0-\sqrt\kappa r_0) \over|a_k|^2+\lambda
   r_0^2}\right)^2  }  
  +\sum_{q\ne 0}{\ts {1\over\kappa}}\sum_k \ts 
  {\lambda \bar\gamma \over|a_k|^2+\lambda r_0^2}
  {\lambda \bar\gamma \over|a_{k-q}|^2+\lambda r_0^2}\,\phi_q
 \,\phi_{-q}  \cr
 &\phantom{=}+  {\ts {1\over\kappa}} \sum_k {\ts
  \left({\lambda r_0 (\rho_0-\sqrt\kappa r_0) \over|a_k|^2+\lambda
   r_0^2}\right)^2  }
  +\sum_{q\ne 0}{\ts {1\over\kappa}}\sum_k \ts 
  {\lambda \gamma \over|a_k|^2+\lambda r_0^2}
  {\lambda \gamma \over|a_{k-q}|^2+\lambda r_0^2}\,\bar\phi_q
  \,\bar\phi_{-q} &(51) \cr
 &\phantom{=}  {\ts -{\lambda\over\kappa}}\sum_k{\ts {  |a_k|^2  
     \over(|a_k|^2+\lambda r_0^2)^2} }(\rho_0-\sqrt\kappa r_0)^2
 - \sum_{q\ne 0}{\ts {\lambda\over\kappa}}\sum_k 
   \ts {  \bar a_k \over|a_k|^2+\lambda r_0^2}\,
  {   a_{k+q} \over|a_{k+q}|^2+\lambda r_0^2}\, \rho_q^2 \cr
 &\phantom{=} {\ts -{\lambda\over\kappa}}\sum_k {\ts {  |a_k|^2  
     \over(|a_k|^2+\lambda r_0^2)^2} }(\rho_0-\sqrt\kappa r_0)^2
 - \sum_{q\ne 0}{\ts {\lambda\over\kappa}}\sum_k
   \ts {  a_k \over|a_k|^2+\lambda r_0^2}\,
  {  \bar a_{k-q} \over|a_{k-q}|^2+\lambda r_0^2}\, \rho_q^2 \biggr\} 
 +O(\xi^3)
 \cr}$$
Using the BCS equation (5), 
   ${\lambda\over\kappa}\sum_k{1\over |a_k|^2+
  \lambda r_0^2}=1$ and abbreviating 
$$E_k^2:=|a_k|^2+\lambda r_0^2=k_0^2+e_\k^2+\lambda r_0^2$$
 this becomes 
$$\eqalignno{2\sqrt\kappa & r_0(\rho_0-\sqrt\kappa r_0)\cdot 1
  -(\rho_0-\sqrt\kappa r_0)^2 {\ts{\lambda\over\kappa}} 
  \sum_k {\ts{\lambda r_0^2-|a_k|^2\over E_k^4}}+
  \sum_{q\ne 0} \rho_q^2 \biggl\{ {\ts{\lambda\over\kappa}}
  \sum_k \ts { \bar a_k a_{k-q} \over E_k^2 E_{k-q}^2}
   \biggr\}  \cr 
  &\phantom{=}-\sum_{q\ne 0}{\rm Re}
    \left( e^{-2i\theta_0}\phi_q\phi_{-q}\right) 
   \biggl\{ {\ts{\lambda\over\kappa}}\sum_k \ts {\lambda r_0^2 
  \over E_k^2 E_{k-q}^2} \biggr\}  \cr
 &=2\sqrt\kappa  r_0(\rho_0-\sqrt\kappa r_0)\cdot 1
  +(\rho_0-\sqrt\kappa r_0)^2-(\rho_0-\sqrt\kappa r_0)^2
  2 {\ts{\lambda\over\kappa}} 
  \sum_k {\ts{\lambda r_0^2 \over E_k^4}} \cr
 &\phantom{=} +
  \sum_{q\ne 0} \rho_q^2 \biggl\{ {\ts{\lambda\over\kappa}}
  \sum_k {\ts { \bar a_k a_{k-q} \over E_k^2 E_{k-q}^2}
   \biggr\} }
  -\sum_{q\ne 0}{\rm Re}\left( e^{-2i\theta_0}\phi_q\phi_{-q}\right) 
   \biggl\{ {\ts{\lambda\over\kappa}}\sum_k \ts {\lambda r_0^2 
  \over E_k^2 E_{k-q}^2} \biggr\}  \cr
  &=\rho_0^2 -\kappa r_0^2-(\rho_0-\sqrt\kappa r_0)^2
  2 {\ts{\lambda\over\kappa}} 
  \sum_k {\ts{\lambda r_0^2 \over E_k^4}} \cr
 &\phantom{=} +
  \sum_{q\ne 0} \rho_q^2 \biggl\{ {\ts{\lambda\over\kappa}}
  \sum_k {\ts { \bar a_k a_{k-q} \over E_k^2 E_{k-q}^2}
   \biggr\} }
  -\sum_{q\ne 0}{\rm Re}\left( e^{-2i\theta_0}\phi_q\phi_{-q}\right) 
   \biggl\{ {\ts{\lambda\over\kappa}}\sum_k \ts {\lambda r_0^2 
  \over E_k^2 E_{k-q}^2} \biggr\} &(52) \cr}$$
Therefore one obtains, recalling that 
 $\xi_{k,p}={1\over\sqrt\kappa}\,\phi_{k-p}-\gamma\,\delta_{k,p}$,  
$$ \eqalignno{ V(\{\phi_q\})&- 
 V(\{\sqrt\kappa \, \delta_{q,0}\, r_0 e^{i\theta_0} \})
    =\sum_q \rho_q^2 -\kappa\, r_0^2\>-\>
  \log\det\left[ Id+ \left(\matrix{ \lambda
 {\bar\gamma \over|a|^2+\lambda r_0^2}\,\xi &
    ig {\bar A \over|a|^2+\lambda r_0^2}\,\xi^* \cr 
   ig {A \over|a|^2+\lambda r_0^2}\,\xi &
    \lambda {\gamma \over|a|^2+\lambda r_0^2}\,\xi^* \cr}
   \right)\right]  \cr 
 &=\sum_{q\ne 0} \rho_q^2 +(\rho_0-\sqrt\kappa r_0)^2
  2 {\ts{\lambda\over\kappa}} 
  \sum_k {\ts{\lambda r_0^2 \over E_k^4}}  -
  \sum_{q\ne 0} \rho_q^2 \biggl\{ {\ts{\lambda\over\kappa}}
  \sum_k {\ts { \bar a_k a_{k-q}\over E_k^2 E_{k-q}^2}
   \biggr\} }\cr
 &\phantom{=}
  +\sum_{q\ne 0}{\rm Re}\left( e^{-2i\theta_0}\phi_q\phi_{-q}\right) 
   \biggl\{ {\ts{\lambda\over\kappa}}\sum_k \ts {\lambda r_0^2 
  \over E_k^2 E_{k-q}^2} \biggr\} &(53) \cr}$$
%
%
%
%
%
Consider the coefficient of $\sum_{q\ne 0}\rho_q^2$. It is given by 
$$\eqalignno{ 1&-{\ts{\lambda\over\kappa}}
  \sum_k {\ts { \bar a_k a_{k-q} \over E_k^2 E_{k-q}^2}}=
 {\ts {1\over2}} (1+1)-{\ts {1\over2}}{\ts{\lambda\over\kappa}}
  \sum_k {\ts { 2\bar a_k a_{k-q}
   \over E_k^2 E_{k-q}^2}}  \cr
 &= {\ts {1\over2}}\biggl({\ts{\lambda\over\kappa}}
     \sum_k{\ts {|a_k|^2+\lambda r_0^2\over 
   E_k^2 E_{k-q}^2}}+  
  {\ts{\lambda\over\kappa}} \sum_k{\ts {|a_{k-q}|^2+\lambda r_0^2\over 
   E_k^2 E_{k-q}^2}}\biggr) -{\ts {1\over2}}{\ts{\lambda\over\kappa}}
  \sum_k {\ts { 2\bar a_k a_{k-q}
   \over E_k^2 E_{k-q}^2}}  \cr
  &= {\ts {1\over2}}{\ts{\lambda\over\kappa}}\sum_k 
  {\ts {a_k\bar a_k-a_k\bar a_{k-q}-\bar a_k a_{k-q}+a_{k-q} 
  \bar a_{k-q}\over E_k^2 E_{k-q}^2}}+ 
  {\ts{\lambda\over\kappa}}
     \sum_k{\ts { \lambda r_0^2\over 
   E_k^2 E_{k-q}^2}}  -{\ts {1\over2}{\lambda\over\kappa} } 
  \sum_k\ts  { \bar a_k a_{k-q} -a_k\bar a_{k-q} \over E_k^2 E_{k-q}^2}     \cr
 &= {\ts {1\over2}}{\ts{\lambda\over\kappa}}\sum_k 
  {\ts {(a_k-a_{k-q})(\bar a_k-\bar a_{k-q})\over E_k^2 E_{k-q}^2}}+ 
  {\ts{\lambda\over\kappa}}
     \sum_k{\ts { \lambda r_0^2\over 
   E_k^2 E_{k-q}^2}}-i{\ts {\lambda\over\kappa} } 
  \sum_k\ts  { {\rm Im}( \bar a_k a_{k-q})  \over E_k^2 E_{k-q}^2}    \cr
  &= {\ts {1\over2}}{\ts{\lambda\over\kappa}}\sum_k 
  {\ts { q_0^2+(e_\k-e_{\k-\q})^2 \over E_k^2 E_{k-q}^2}}+ 
  {\ts{\lambda\over\kappa}}
     \sum_k{\ts { \lambda r_0^2\over 
   E_k^2 E_{k-q}^2}}
 -i{\ts {\lambda\over\kappa} } 
  \sum_k\ts {  k_0e_{\k-\q}-(k_0-q_0)e_\k  \over E_k^2 E_{k-q}^2}  \cr
 &= \alpha_q+i\gamma_q+\beta_q   &(54) \cr}$$ 
Inserting (54) in (53), one gets 
$$ \eqalignno{ V(\{\phi_q\})&- 
 V(\{\sqrt\kappa \, \delta_{q,0}\, r_0 e^{i\theta_0} \})=
    (\rho_0-\sqrt\kappa r_0)^2
  2 \beta_0  +
  \sum_{q\ne 0} \rho_q^2 \biggl\{1- {\ts{\lambda\over\kappa}}
  \sum_k {\ts { 2\bar a_k a_{k-q}\over E_k^2 E_{k-q}^2}
   \biggr\} }\cr
 &\phantom{=}
  +\sum_{q\ne 0}{\rm Re}\left( e^{-2i\theta_0}\phi_q\phi_{-q}\right) 
   \biggl\{ {\ts{\lambda\over\kappa}}\sum_k \ts {\lambda r_0^2 
  \over E_k^2 E_{k-q}^2} \biggr\}  \cr
 &= (\rho_0-\sqrt\kappa r_0)^2
  2 \beta_0 +
  \sum_{q\ne 0} \rho_q^2 ( \alpha_q+i\gamma_q)  \cr 
  &\phantom{=}+\sum_{q\ne 0} \rho_q^2 \beta_q
  +\sum_{q\ne 0}{\ts { e^{-2i\theta_0}\phi_q\phi_{-q}+
   e^{2i\theta_0}\bar\phi_q\bar\phi_{-q}\over2}} \beta_q &(55)  \cr}$$
Since $\beta_q=\beta_{-q}$,   
the last two q-sums in (55) may be combined to give 
$$\eqalignno{ \sum_{q\ne 0}& {\ts {\rho_q^2+\rho_{-q}^2\over 2}}
  \beta_q 
  +\sum_{q\ne 0}{\ts { e^{-2i\theta_0}\phi_q\phi_{-q}+
   e^{2i\theta_0}\bar\phi_q\bar\phi_{-q}\over2}}
    \beta_q  \cr
 &={\ts{1\over2}}\sum_{q\ne 0}( \phi_q\bar\phi_q 
  +\phi_{-q}\bar\phi_{-q}+e^{-2i\theta_0}\phi_q\phi_{-q}+
   e^{2i\theta_0}\bar\phi_q\bar\phi_{-q})
   \beta_q \cr
 &={\ts{1\over2}}\sum_{q\ne 0}(e^{-i\theta_0} \phi_q
   +e^{i\theta_0} \bar\phi_{-q})
    (e^{i\theta_0} \bar\phi_q+e^{-i\theta_0}\phi_{-q})
    \beta_q\cr
  &={\ts{1\over2}}\sum_{q\ne 0}|e^{-i\theta_0} \phi_q
   +e^{i\theta_0} \bar\phi_{-q}|^2 
    \beta_q  &(56)
    \cr}$$
This proves Theorem 2 $\blacksquare$
\bigskip
\bigskip
\noindent{\gross III. The Effective Potential with a U(1)
 Symmetry Breaking}
\par\noindent{\gross$\phantom{Inn. }$ External Field}
\bigskip
We consider now the situation where a small external field is 
added to the action which breaks the $U(1)$ symmetry. In that case, 
the partition function (13) changes to 
$$Z_r=\int e^{ {\lambda\over(\kappa)^3}\sum_{k,p,q}\bar\psi_{k\up}
    \bar\psi_{q-k\down}\psi_{q-p\down}\psi_{p\up}
  +{1\over \kappa}\sum_k [r\psi_{k\up}\psi_{-k\down}
  +\bar r\bar\psi_{-k\down}\bar\psi_{k\up}] } d\mu_C\eqno (57)$$
After a Hubbard-Stratonovich transformation, this becomes 
$$Z_r=\int e^{-V_r(\{\phi_q\})} \pro_q 
         {\ts {d\phi_qd\bar\phi_q\over \pi}} \eqno (58)$$
where (recall that $\kappa:=\beta L^d$)   
$$\eqalignno{ V_r(\{\phi_q\})&=\sum_q |\phi_q|^2
  -\log{ \det\left[\matrix{ Id &  
 C ({ig\phi^*\over \sqrt{\kappa}}-\bar r\delta_{k,p}) \cr 
   \bar C({ig\phi\over \sqrt{\kappa}} +r\delta_{k,p}) & 
  Id \cr}\right]  } & (59) \cr}$$
For the, say, $\la \bar\psi_\sigma \psi_\sigma\ra$ and 
 $\la\psi_\up\psi_\down\ra$ correlations one obtains similarly [5]:
$$\eqalignno{ 
  \la \bar\psi_{k\up}\psi_{k\up}\ra_r&=\kappa \la F_r(k) \ra_r
 &(60) \cr
  \la \psi_{k\up}\psi_{-k\down}\ra_r&=\kappa \la G_r(k) \ra_r 
  &(61)\cr}$$
where 
$$\eqalignno{ F_r(k)=F_r(k;\phi)&
   =\left[\left(\matrix{ a_s\delta_{s,t} & 
  ( {ig\bar\phi_{t-s}\over \sqrt{\kappa}} 
  -\bar r\delta_{s,t}) \cr
    ({ig\phi_{s-t}\over \sqrt{\kappa}} +r\delta_{s,t}) & 
  a_{-s}\delta_{s,t} \cr} \right)_{s,t}\right]^{-1}_{k\up,k\up}
  &(62)   \cr
    G_r(k)=G_r(k;\phi)&
   =\left[\left(\matrix{ a_s\delta_{s,t} & 
 ({ig\bar\phi_{t-s}\over \sqrt{\kappa}}-\bar r\delta_{s,t})
   \cr
    ({ig\phi_{s-t}\over \sqrt{\kappa}}+r\delta_{s,t}) & 
  a_{-s}\delta_{s,t} \cr} \right)_{s,t}\right]^{-1}_{k\down,k\up}
   &(63)   \cr}$$
and in (60,61) the expectation on the left is given by the Grassmann
integral with external field and the expectation on the right is given 
by $\la F\ra_r=\int F(\phi)\, e^{-V_r(\phi)}/\int e^{-V_r(\phi)}$.
\par
In (59-63), the external field $r$ only shows up in conjunction with the 
 $\phi_0$ variable through the combination
${\phi_0\over\sqrt{\kappa}}-i{r\over g}$. By 
substitution of variables one has, if $\phi_0=u_0+iv_0$ and 
 $r=|r|\,e^{i\alpha}$ 
$$\int_{\Bbb R^2}{\ts f\Bigl( {\phi_0\over\sqrt{\kappa}}
  -i{r\over g},{\bar\phi_0\over\sqrt{\kappa}}
   +i{\bar r\over g}\Bigr) } e^{-(u_0^2+v_0^2)} du_0dv_0=
  \eqno (64)$$
$$\int_{\Bbb R^2}\ts f\Bigl(e^{i\alpha}
  {\phi_0\over\sqrt{\kappa}},
   e^{-i\alpha}{\bar\phi_0\over\sqrt{\kappa}}\Bigr)\, 
  e^{-\left( u_0^2+(v_0+\sqrt{\kappa}\, {|r|\over g})^2\right)}
    du_0dv_0$$
Thus we can write 
$$Z_r=\int e^{-U_r(\{\phi_q\})} \pro_q 
         {\ts {d\phi_qd\bar\phi_q\over \pi}} \eqno (65)$$
where 
$$\eqalignno{ U_r(\{\phi_q\})&=u_0^2+{\ts
   \left(v_0+\sqrt{\kappa}\,{|r|\over g}\right)^2}
   +\sum_{q\ne 0} |\phi_q|^2
  -\log{ \det\left[\matrix{ Id & {ig\over 
  \sqrt{\kappa} }\, 
    C \tilde\phi^* \cr {ig\over\sqrt{\kappa} }\, 
   \bar C\tilde\phi & Id \cr}\right]  } & (66) \cr}$$
and
$$\tilde \phi=\left( \tilde\phi_{k-p}\right)_{k,p},\;\;\;\;\; 
\tilde\phi_q=\cases{ \phi_q &if $q\ne 0$ \cr e^{i\alpha}\phi_0 & 
  if $q=0$. \cr} \eqno (67)$$
Furthermore 
$$\eqalignno{ 
  \la \bar\psi_{k\up}\psi_{k\up}\ra_r&=\kappa
    \la \tilde F_0(k) \ra_{U_r} &(68) \cr
  \la \psi_{k\up}\psi_{-k\down}\ra_r&=\kappa 
   \la \tilde G_0(k) \ra_{U_r} & (69)
   \cr}$$
where $\tilde F_0(k)$ and $\tilde G_0(k)$ are given by (62,63) with 
  $r=0$ and $\phi$ substituted by $\tilde \phi$. The 
 expectations on the right hand side of (68,69) are now taken 
 with respect 
 to $U_r$, that is $\la F\ra_{U_r}=\int F(\phi)\, e^{-U_r(\phi)} / 
  \int  e^{-U_r(\phi)}$. 
\smallskip
Thus in the case with a small external field we would ask for the 
global minimum of $U_r$ and for the second order Taylor expansion 
 around it. One has the following 
\bigskip
\noindent{\bf Corollary:} {\it  Let $U_r$ be the effective potential 
 (66) with a small external U(1) symmetry breaking field 
 $r=|r|\,e^{i\alpha}$. Let $\tilde\phi$ be given by (67).  Then:
\item{(i)}
The global minimum of 
$${\rm Re}U_r(\{\phi_q\})=u_0^2+{\ts
   \left(v_0+\sqrt{\kappa}\,{|r|\over g}\right)^2}
   +\sum_{q\ne 0} |\phi_q|^2
  -\log{ \left|\det\left[\matrix{ Id & {ig\over 
  \sqrt{\kappa} }\, 
    C \tilde\phi^* \cr {ig\over\sqrt{\kappa} }\, 
   \bar C\tilde\phi & Id \cr}\right]\right|  } $$
is unique and is given by 
 $\phi_q^{\rm min}=\delta_{q,0}\sqrt{\kappa}
\,i y_0$ where $y_0=y_0(|r|)$ 
is the unique global minimum of the function 
  $V_{{\rm BCS},r}:\Bbb R\to 
 \Bbb R$, 
$$\eqalignno{ V_{{\rm BCS},r}(y)&:=
  U_r\left(u_0=0,\,v_0=\sqrt{\kappa}\,y;\>
  \phi_q=0\,\,{\rm for}\,\,q\ne0\right) \cr
 &=\kappa\biggl\{ {\ts \left(y+{|r|\over g}\right)^2} 
  -{\ts {1\over \kappa}}\sum_k\ts 
    \log\left[ 1+{\lambda y^2\over 
    k_0^2+e_\k^2}\right] \biggr\}. &(70)\cr}$$
\item{(ii)}  The second order Taylor expansion of $U_r$ 
 around $\phi^{\rm min}$ is given by  
$$\eqalignno{ U_r\bigl(\{\phi_q\}\bigr)&=U_{r,{\rm min}}
  +2\beta_0(v_0-\sqrt{\kappa}y_0)^2+\sum_{q\ne 0} 
  (\alpha_q+i\gamma_q) |\phi_q|^2
  +{\ts {1\over2}} 
  \sum_{q\ne 0}\beta_q |e^{-i\alpha} \phi_q -e^{i\alpha}\bar\phi_{-q} 
   |^2  \cr
 & \phantom{mm}+{\ts {|r|\over g|y_0|}}\Bigl( u_0^2+(v_0- 
  \sqrt{\kappa} y_0)^2
  +\sum_{q\ne 0} |\phi_q|^2 \Bigr)
   +O\bigl( (\phi-\phi^{\rm min})^3\bigr) . &(71) \cr}$$
where $U_{r,{\rm min}}:=U_r\bigl(\{\phi_q^{\rm min}\})$ and 
 the coefficients $\alpha_q$, $\beta_q$ and $\gamma_q$ are 
 given by (42) of Theorem 2 but $E_k$ in this case is given by 
 $E_k^2=|a_k|^2+\lambda y_0^2=k_0^2+e_\k^2+\lambda y_0^2$.   }
\bigskip
\noindent{\bf Remark:} Of course one has $\lim_{|r|\to 0} 
  \lambda y_0(|r|)^2=\lambda r_0^2=\Delta^2$ 
 where $\pm r_0$ is the global minimum of  $V_{{\rm BCS},r=0}$. 
\bigskip
\noindent{\bf Proof:} {\bf (i)} As in 
 the proof of Theorem I one shows that 
$$\eqalignno{ \log{ \left|\det\left[\matrix{ Id & {ig\over 
  \sqrt{\kappa} }\, 
    C \tilde\phi^* \cr {ig\over\sqrt{\kappa} }\, 
   \bar C\tilde\phi & Id \cr}\right]\right|  } &\le 
  \sum_k \ts \log\left[ 1+{ {\lambda\over \kappa}\sum_q|\phi_q|^2 
  \over |a_k|^2 } \right] \cr
 &=: \sum_k \ts \log\left[ 1+{\lambda(x^2+y^2)  
  \over |a_k|^2 } \right]  &(72)\cr}$$
where we abbreviated 
$$x^2:= {\ts {1\over \kappa}} \Bigl( u_0^2+\sum_{q\ne 0}|\phi_q|^2 
  \Bigr),\;\;\;\; y^2:={\ts {1\over \kappa}} v_0^2\eqno (73)$$
Thus 
$$\eqalignno{ {\rm Re}U_r(\{\phi_q\})&\ge {\ts\bigl( v_0+\sqrt\kappa 
  {|r|\over g}\bigr)^2}+u_0^2+\sum_{q\ne 0}|\phi_q|^2 
  -\sum_k \ts \log\left[ 1+{\lambda(x^2+y^2)  
  \over |a_k|^2 } \right] \cr 
 &=:\kappa W_r(x,y) &(74) \cr}$$
where 
$$W_r(x,y)= x^2+{\ts\bigl( y+ {|r|\over g}\bigr)^2}
  -{\ts {1\over \kappa}} 
  \sum_k \ts \log\left[ 1+{\lambda(x^2+y^2)  
  \over |a_k|^2 } \right] \eqno (75) $$
The global minimum of $W_r$ is unique and given by $x=0$ and 
 $y=y_0$ where $y_0$ is the unique global minimum of (70). Since 
$ U_r\left(u_0=0,\,v_0=\sqrt{\kappa }\,y;\>
  \phi_q=0\,\,{\rm for}\,\,q\ne0\right) =V_{{\rm BCS},r}(y)$, 
part (i) follows. 
\medskip
\noindent {\bf (ii)} Part (ii) is proven in the same way 
 as Theorem II. One has 
$$\eqalignno{ U_r(\{\phi_q\})&-U_{r,{\rm min}}= 
 u_0^2+\sum_{q\ne 0}|\phi_q|^2 +\bigl( v_0+\sqrt\kappa
  {\ts {|r|\over g}}\bigr)^2- \bigl( \sqrt\kappa\, y_0+\sqrt\kappa
  {\ts {|r|\over g}}\bigr)^2  \cr 
 &\phantom{n} -\log\Biggl\{ \det
  \left[\matrix{ a_k\delta_{k,p} & 
   {ig\over \sqrt\kappa }\bar{\tilde\phi}_{p-k}  \cr
     {ig\over \sqrt\kappa }\tilde\phi_{k-p} & 
  a_{-k}\delta_{k,p} \cr} \right] \Bigr/ \det 
   \left[\matrix{ a_k\delta_{k,p} & 
   {ig\over \sqrt\kappa }e^{-i\alpha}(-i)y_0\delta_{k,p} \cr
     {ig\over \sqrt\kappa} e^{i\alpha}iy_0\delta_{k,p} & 
  a_{-k}\delta_{k,p} \cr} \right] \Biggr\} \cr
 & \cr  
 &=u_0^2+\sum_{q\ne 0}|\phi_q|^2 +(v_0-\sqrt\kappa y_0)^2
  +2(v_0-\sqrt\kappa y_0)\bigl(\sqrt\kappa y_0+
   \sqrt\kappa{\ts{|r|\over g}}\bigr) &(76) \cr
 &\phantom{n} -\log\det\biggl[ Id+
  \left(\matrix{ a_k\delta_{k,p} & 
   {ig\over \sqrt\kappa }e^{-i\alpha}(-i)y_0\delta_{k,p} \cr
     {ig\over \sqrt\kappa} e^{i\alpha}iy_0\delta_{k,p} & 
  a_{-k}\delta_{k,p} \cr} \right)^{-1}
  \left(\matrix{ 0 & 
   {ig\over \sqrt\kappa }\bar\xi_{p-k} \cr
     {ig\over \sqrt\kappa }\xi_{k-p}
   &  0 \cr} \right)\biggr]  \cr}$$
  where in this case 
$$\xi_{k-p}:=\tilde\phi_{k-p}-\sqrt\kappa
   e^{i\alpha}iy_0\delta_{k,p} \eqno (77)$$
The expression $\log\det[Id+\cdots]$ is expanded as in the proof of 
 Theorem II. One obtains, if $E_k^2:=|a_k|^2+\lambda y_0^2$, 
$$\eqalignno{ \log\det[Id+\cdots]&=2\sqrt\kappa y_0 
  (v_0-\sqrt\kappa y_0) {\ts {\lambda\over \kappa}} 
  \sum_k{\ts {1\over E_k^2}} \cr 
 &\phantom{mm} +{\ts {\lambda\over \kappa}} 
  \sum_k {\ts {\lambda y_0^2 \left( u_0^2-(v_0-\sqrt\kappa y_0)^2
   \right) \over E_k^4} } +{\ts {\lambda\over \kappa}} \sum_k
  { \ts{ |a_k|^2 \left( u_0^2+(v_0-\sqrt\kappa y_0)^2
   \right) \over E_k^4} } \cr
 &\phantom{mm} + {\ts {1\over 2}{\lambda\over \kappa}} 
  \sum_{q\ne 0} \sum_k \left( {\ts 
  {a_k\bar a_{k-q}\over E_k^2 E_{k-q}^2} + 
  {\bar a_k a_{k+q}\over E_k^2 E_{k+q}^2}  }\right)|\phi_q|^2 \cr
 &\phantom{mm} - {\ts {1\over 2}{\lambda\over \kappa}} 
  \sum_{q\ne 0} \sum_k \left( {\ts {\lambda y_0^2 e^{2i\alpha} 
  \over E_k^2 E_{k+q}^2}  \bar\phi_q\bar\phi_{-q}
  + {\lambda y_0^2 e^{-2i\alpha} 
  \over E_k^2 E_{k-q}^2} \phi_q\phi_{-q} } \right)
   &(78)  \cr}$$
Since $y_0$ is a minimum of $V_{{\rm BCS},r}$, one has the BCS 
equation 
$$2\left( y_0+{\ts {|r|\over g}}\right) -{\ts {\lambda\over \kappa}} 
  \sum_k {\ts {2y_0\over E_k^2}} =0 \;\;\;{\rm or}\;\;\;
    {\ts {\lambda\over \kappa}} 
  \sum_k \ts {1\over E_k^2} =1 -\ts {|r|\over g|y_0|}  \eqno (79)$$
Using this, one gets (observe that $y_0$ is negative) 
$$\eqalignno{ U_r(\{\phi_q\})&=U_{r,{\rm min}}
  +{\ts {|r|\over g|y_0|}} \left( 
  u_0^2+(v_0-\sqrt\kappa\, y_0)^2\right)+2{\ts {\lambda\over\kappa}}
  \sum_k {\ts {\lambda y_0^2(v_0-\sqrt\kappa y_0)^2\over E_k^4}}
  &(80) \cr 
 &\phantom{mm}+  
 \sum_{q\ne 0} \biggl\{ 1-{\ts {\lambda\over \kappa}} 
  \sum_k {\ts {\bar a_k a_{k-q}\over E_k^2 E_{k-q}^2}} \biggr\} 
  |\phi_q|^2 
  +{\ts {1\over 2}{\lambda\over \kappa} } 
  \sum_{q\ne 0} \sum_k \left( {\ts {\lambda y_0^2 e^{2i\alpha} 
  \over E_k^2 E_{k+q}^2}  \bar\phi_q\bar\phi_{-q}
  + {\lambda y_0^2 e^{-2i\alpha} 
  \over E_k^2 E_{k-q}^2} \phi_q\phi_{-q} } \right) \cr}$$
Using the BCS equation (79) again, one obtains (compare (54))
$$1-{\ts {\lambda\over \kappa}} 
  \sum_k {\ts {\bar a_k a_{k-q}\over E_k^2 E_{k-q}^2}}= 
  \alpha_q+i\gamma_q +\beta_q +\ts 
  {|r|\over g|y_0|}$$ 
Substituting this in (80) and rearranging as in the proof of Theorem
II proves part (ii) $\blacksquare$

\bigskip
\bigskip
\bigskip
\noindent{\gross References}
\bigskip
\item{[1]} J. Feldman, J. Magnen, V. Rivasseau, E. Trubowitz, {\it Fermionic 
 Many Body Models}, in: CRM Proceedings and Lecture Notes Vol. 7, 
 {\it Mathematical Quantum Theory I: Field Theory and Many Body Theory}, 
 eds. J. Feldman, R. Froese, L. M. Rosen, 1994.
\item{[2]} J. Feldman, J. Magnen, V. Rivasseau, E. Trubowitz, {\it Ward Identities 
 and a Perturbative Analysis of a U(1) Goldstone Boson in a Many Fermion 
 System}, Helvetia Physica Acta 66, 1993, 498-550.
\item{[3]}  T. Chen,  J. Fr\"ohlich,  M. Seifert, {\it Renormalization Group 
  Methods: Landau-Fermi Liquid and BCS Superconductor}, Proceedings of the 
 Les Houches session {\it Fluctuating Geometries in Statistical Mechanics 
 and Field Theory}, eds. F. David, P. Ginsparg, J. Zinn-Justin, 1994. 
\item{[4]} J. Feldman, E. Trubowitz, {\it Perturbation Theory for Many  
 Fermion Systems}, Helv. Phys. Acta 63, 1990, 156-260; {\it The Flow of 
 an Electron Phonon System to the Superconducting State}, Helv. Phys. 
 Acta 64, 1991, 214-357. 
\item{[5]} D. Lehmann, {\it The Many-Electron System in the Forward, 
 Exchange and BCS Approximation}, Comm. Math.  
Phys. 198, 427-468, 1998.

\end